\documentclass[10pt,english]{article}
\usepackage{mathptmx}
\usepackage[T1]{fontenc}
\usepackage[latin9]{luainputenc}
\usepackage{babel}
\usepackage{mathrsfs}
\usepackage{amsmath}
\usepackage{amsthm}
\usepackage{amssymb}
\usepackage{graphicx}
\usepackage{setspace}
\usepackage[pdfusetitle,
 bookmarks=true,bookmarksnumbered=false,bookmarksopen=false,
 breaklinks=false,pdfborder={0 0 1},backref=false,colorlinks=false]
 {hyperref}
\begin{document}
\title{Blanket representation for spin networks with $n-$body interactions}
\author{Simone Franchini{\normalsize\thanks{Correspondence: simone.franchini@yahoo.it}\thanks{CNR-ISTC, Via Gian Domenico Romagnosi 18, 00196 Rome, Italy}}}
\date{~}
\maketitle
\begin{abstract}
This paper extends the Blanket representation of {[}\textit{Universal
scaling limits for spin networks via martingale methods}, Franchini,
S., Proc. R. Soc. A, \textbf{481} (2025){]} from systems with two--body
interactions to multi--spin (or $n-$body) interactions. This generalization
allows for the exploration of broader physical phenomena where higher--order
interactions are significant.

~

\noindent\textit{keywords: multi--spin networks, Boltzmann machines,
neural networks, many body interactions}
\end{abstract}
\newpage{}

\section{Introduction}

In \cite{RSBwR,Franchini2021,FranchiniSPA2023,Franchini2024,Franchini_Pan_2025}
we introduced a mathematical framework that allows to approach disordered
systems through the lens of Large--Deviations (LD) theory \cite{FranchiniURNS2017,FranchiniBalzanIRT2023,Dembo}.
This method has proven particularly effective for analyzing spin systems,
including the Sherrington--Kirkpatrick (SK) model \cite{RSBwR,Franchini2021,FranchiniSPA2023},
which is central for the celebrated \cite{ParisiNobel} ``Replica
Symmetry Breaking'' (RSB) Theory of Parisi, Mezard, Virasoro \cite{Parisi_Mezard_Virasoro}
and many others \cite{Charbonneau}, and represents a mean--field
paradigm for disordered systems \cite{ParisiNobel} and neural networks
\cite{Amit,Rende}. Here we extended the Large--Deviations theory
of \cite{RSBwR,Franchini2021,FranchiniSPA2023,Franchini2024,Franchini_Pan_2025}
from networks characterized by two--body interactions to more complex
systems involving multi--spin ($n-$body) interactions \cite{Battiston}.
This generalization is crucial one, as it opens to the exploration
of much broader spectrum of physical phenomena where the interactions
beyond pairwise are significant \cite{Battiston,Battiston_long,BFPF,Bardella} 

\subsection{The Model}

Hereafter we assume the notation of \cite{Franchini2024}. Let name
the vertex set with
\begin{equation}
V:=\left\{ 1\leq i\leq N\right\} 
\end{equation}
and denote the spin state with the symbol
\begin{equation}
\sigma_{V}:=\left\{ \sigma_{1},\,...\,,\,\sigma_{N}\right\} \in\left\{ -1,1\right\} ^{V}
\end{equation}
as in \cite{Franchini2024}. In what follows we adopt the convention
that 
\begin{equation}
W^{(n)}:=V^{n}
\end{equation}
Summing over this set is equivalent to sum on the coordinates
\begin{equation}
\sum_{i_{1}\in V}\ ...\ \sum_{i_{n}\in V}=\sum_{i_{1}\,...\,i_{n}\in W^{(n)}}
\end{equation}
In general, a spin Hamiltonian with $n-$body interactions can be
controlled by a tensor of parameters with at least $n$ dimensions
\begin{equation}
H^{\left(n\right)}:\left\{ -1,1\right\} ^{V^{n}}\rightarrow\mathbb{R}
\end{equation}
that can be random. From this tensor the Hamiltonian is explicitly
written as follows:
\begin{equation}
H^{\left(n\right)}\left(\sigma_{V}\right):=\sum_{i_{1}...\,i_{n}\in W}H_{i_{1}\,...\,i_{n}}^{\left(n\right)}\sigma_{i_{1}}\,...\,\sigma_{i_{n}}
\end{equation}
The canonical partition function and the associated Gibbs probability
measure are defined through the following formulas:
\begin{equation}
Z:=\sum_{\sigma_{V}\in\left\{ -1,1\right\} ^{V}}\exp\,H^{\left(n\right)}\left(\sigma_{V}\right),\ \ \ \mu^{*}\left(\sigma_{V}\right):=\frac{\exp\,H^{\left(n\right)}\left(\sigma_{V}\right)}{Z}
\end{equation}
In the present paper our interest will be especially focused in finding
a general formula for the pressure density per spin (proportional
to the free energy):
\begin{equation}
\mathbb{E}f:=\frac{1}{N}\,\mathbb{E}\log Z
\end{equation}
where we indicated with the notation $\mathbb{E}$ the average respect
to the possible randomness of the parameter matrix. 

\section{Free Energy principle}

We will now recall the general findings from \cite{RSBwR,Franchini2021,FranchiniSPA2023,Franchini2024,Franchini_Pan_2025}
on the quenched free energy. In particular, we will sketch how to
obtain a generalized analogue \cite{Franchini2024} of the Parisi--type
\cite{Guerra_Bound,ASS,From_Parisi_to_Boltzmann} variational bounds
for the quenched free energy by partitioning the systems into a progression
of Markov blankets \cite{Pearl,Friston}. We follow the formulation
in \cite{Franchini2024}.

\subsection{The order parameter}

Following \cite{Franchini2024} we introduce the re--normalized vertex
set
\begin{equation}
\Lambda:=\left\{ 1\leq\ell\leq L\right\} 
\end{equation}
with $L\leq N$ vertices. Hereafter we assume that $L$ is a finite
integer independent of $N$ unless specified otherwise. Following
\cite{Franchini2024} we identify the order parameter in a cumulative
distribution function (CDF) \cite{Steinbrecher} with $L$ atoms:
\begin{equation}
q_{\Lambda}:=\left\{ q_{1},\,...\,,\,q_{L}\right\} \in\mathcal{P}\left(\Lambda\right)
\end{equation}
As shown in \cite{RSBwR} is equivalent to the overlap of the RSB
theory \cite{Parisi_Mezard_Virasoro,Charbonneau}. The support of
the order parameter is therefore the space of CDFs with $L$ atoms
\begin{equation}
\mathcal{P}\left(\Lambda\right):=\{q_{\Lambda}\in\left[0,1\right]^{\,\Lambda}:\,q_{\ell}\geq q_{\ell-1}\}
\end{equation}
Following the notation of \cite{Franchini2024}, the theory can be
better exposed by introducing a notation also for the actual atoms
of the distribution, i.e., 
\begin{equation}
p_{\ell}:=q_{\ell}-q_{\ell-1}
\end{equation}
Using this notation we can rewrite the order parameter in terms of
the atomic probability mass function (PMF), that we arrange in the
vector
\begin{equation}
p_{\Lambda}:=\left\{ p_{1},\,...\,,\,p_{L}\right\} \in\mathcal{P}\left(\Lambda\right)
\end{equation}
Since each $p_{\Lambda}$ is associated to only one $q_{\Lambda}$,
the PMF support 
\begin{equation}
\mathcal{P}\left(\Lambda\right)=\{p_{\Lambda}\in\left[0,1\right]^{\,\Lambda}:\,1_{\Lambda}\cdot p_{\Lambda}=1\}
\end{equation}
is isomorphic to the previous one, formulated in terms of CDFs. In
what follows we will use both definitions at our convenience.

\subsection{The Blanket partition}

We will now sketch the Blanket representation as we described it in
\cite{Franchini2024}. The first step to re--normalize the system
is to apply a partition of the vertex set $V$ into $L$ parts: 
\begin{equation}
V=\left\{ V_{1},\,...\,,\,V_{L}\right\} 
\end{equation}
We used the subscript $\ell$ to track the parts of $V$. Following
the blanket representation of \cite{Franchini2024} we introduce the
sequence of sets
\begin{equation}
Q_{\ell}:=\left\{ V_{1},\,...\,,\,V_{\ell}\right\} ,\ \ \ Q_{L}=V
\end{equation}
that is obtained by joining the parts of $V$ up to a certain level
$\ell$. We indicated the spin fields associated to those vertex sets
with the following symbols:
\begin{equation}
\sigma_{Q_{\ell}}:=\left\{ \sigma_{V_{1}},\,...\,,\,\sigma_{V_{\ell}}\right\} ,\ \ \ \sigma_{V}=\sigma_{Q_{L}}=\left\{ \sigma_{V_{1}},\,...\,,\,\sigma_{V_{L}}\right\} 
\end{equation}
The size of the parts are controlled by the order parameter
\begin{equation}
\left|V_{\ell}\right|/N=p_{\ell},\ \ \ \left|Q_{\ell}\right|/N=q_{\ell}
\end{equation}
That given, the edges set supporting the $\ell-$th blanket is defined
as the subset of edges with all ends in $Q_{\ell}^{n}$ minus those
with all ends in $Q_{\ell-1}^{n}$, i.e.,
\begin{equation}
W_{\ell}^{(n)}:=Q_{\ell}^{n}/Q_{\ell-1}^{n}
\end{equation}
The size (cardinality) of the blanket is linked to the order parameter
by
\begin{equation}
|W_{\ell}^{(n)}|/N^{n}=q_{\ell}^{n}-q_{\ell-1}^{n}
\end{equation}
That is in fact the $n-$body generalization of the correction term
appearing in the Parisi functional of the SK model ($n=2$). 

\subsection{Lattice Field Theory}

We recall the Lagrangian formalism described in \cite{RSBwR,Franchini2021,FranchiniSPA2023,Franchini2024},
that builds on findings by Lovász \cite{Lovasz} Borgs, Chayes \cite{Borgs2,Borgs3}
Coja--Oghlan \cite{Coja-Oghlan} and many others, and involves partitioning
the graph into a progression of Markov blankets, \cite{Pearl,Friston}.
We adopted this method for the first time in \cite{FranchiniMS2011}
in the context of polymer physics, see also \cite{FranchiniURNS2017,FranchiniBalzanIRT2023,FranchiniPhD2015,FranchiniBalzanRANGE2018}.
Let introduce the Lagrangian function
\begin{equation}
\mathcal{L}_{\ell}:\left\{ -1,1\right\} ^{Q_{\ell}}\rightarrow\mathbb{R}
\end{equation}
that is obtained from the full Hamiltonian as follows:
\begin{equation}
\mathcal{L}_{\ell}\left(\sigma_{Q_{\ell}}\right):=H^{\left(n\right)}\left(\sigma_{Q_{\ell}}\right)-H^{\left(n\right)}\left(\sigma_{Q_{\ell-1}}\right)\label{eq:LAGRANGIAN}
\end{equation}
and that we interpret as the Hamiltonian associated to the $\ell-$th
Blanket. The Hamiltonian of the full system is then recovered by rejoining
the blankets 
\begin{equation}
H^{\left(n\right)}\left(\sigma_{V}\right)=\sum_{\ell\in\Lambda}\mathcal{L}_{\ell}\left(\sigma_{Q_{\ell}}\right)\label{eq:HAMILTONIAN_ACTION}
\end{equation}
Notice that $\mathcal{L}_{\ell}$ depends only on the spins of $Q_{\ell}$,
that is why we are allowed to interpret the Hamiltonian as an Action
function, and the blanket index $\ell$ as a fictional time. We could
see this as the action of a perturbation propagating trough the graph.
Following \cite{RSBwR,Franchini2021,FranchiniSPA2023,Franchini2024}
we introduce the Gibbs measure of the blanket
\begin{equation}
\xi_{\ell}\left(\sigma_{Q_{\ell}}\right):=\frac{\exp\,\mathcal{L}_{\ell}\left(\sigma_{Q_{\ell}}\right)}{Z_{\ell}\left(\sigma_{Q_{\ell-1}}\right)}\label{eq:LAYER_measure}
\end{equation}
where the normalization is the fundamental quantity
\begin{equation}
Z_{\ell}\left(\sigma_{Q_{\ell-1}}\right):=\sum_{\sigma_{\,V_{\ell}}\in\left\{ -1,1\right\} ^{V_{\ell}}}\exp\,\mathcal{L}_{\ell}\left(\sigma_{Q_{\ell}}\right)\label{eq:LAYER_PF}
\end{equation}
that is, the partition function of the individual blanket. Notice
that it depends not only on the spins of the $\ell-$th blanket, but
also on the previous. The full blanket measure $\mu$ is the product
measure of the blankets:
\begin{equation}
\mu\left(\sigma_{V}\right)=\prod_{\ell\in\Lambda}\xi_{\ell}\left(\sigma_{Q_{\ell}}\right)
\end{equation}
The average of any observable $\mathcal{O}$ is found through the
following recursion: 
\begin{equation}
\mathcal{O}_{\ell-1}\left(\sigma_{Q_{\ell-1}}\right):=\langle\mathcal{O}_{\ell}\left(\sigma_{Q_{\ell}}\right)\rangle_{\xi_{\ell}}\label{eq:therecursion-1-1}
\end{equation}
The starting object is the observable $\mathcal{O}$ itself, then
we apply the averages respect to the blankets until the last one,
that is the average respect to $\mu$:
\begin{equation}
\mathcal{O}_{L}\left(\sigma_{Q_{L}}\right):=\mathcal{O}\left(\sigma_{V}\right),\ \ \ \langle\mathcal{O}\left(\sigma_{V}\right)\rangle_{\mu}=\mathcal{O}_{0}\label{eq:therecursion-1-3}
\end{equation}
Notice that although each blanket depends from all the previous the
average respect to $\xi_{\ell}$ only integrates the spins of that
blanket, i.e., the spins of $V_{\ell}$.

\subsection{Variational bound}

We can finally state the variational bounds for the pressure. Following
\cite{Franchini2024}, let us introduce the blanket pressure, that
is proportional to the logarithm of the partition function
\begin{equation}
f_{\ell}\left(\sigma_{Q_{\ell-1}}\right):=\frac{1}{\left|V_{\ell}\right|}\log Z_{\ell}\left(\sigma_{Q_{\ell-1}}\right)
\end{equation}
summing together we obtain the full pressure 
\begin{equation}
f\left(\sigma_{V}\right):=\langle f_{\ell}\left(\sigma_{Q_{\ell-1}}\right)\rangle_{p}=\sum_{\ell\leq L}p_{\ell}f_{\ell}\left(\sigma_{Q_{\ell-1}}\right)
\end{equation}
This functional is flat respect to the order parameter: 
\begin{equation}
\mathbb{E}f=\frac{1}{N}\,\mathbb{E}\,\log\,\langle\exp\left(Nf\left(\sigma_{V}\right)\right)\rangle_{\mu},\ \ \ \forall p_{\Lambda}\in\mathcal{P}\left(\Lambda\right)
\end{equation}
Introducing the analogue of the Parisi functional
\begin{equation}
\mathcal{G}\left(p_{\Lambda}\right):=\frac{1}{N}\,\mathbb{E}\,\log\,\langle\langle\exp\left(Nf_{\ell}\left(\sigma_{Q_{\ell-1}}\right)\right)\rangle_{\mu}\rangle_{p}
\end{equation}
we deduce from Jensen inequality (ie., bringing the average respect
to the order parameter outside the exp) \cite{Franchini2024} that
it is, remarkably, an upper bound for the pressure:
\begin{equation}
\mathbb{E}f\leq\inf_{p_{\Lambda}\in\mathcal{P}\left(\Lambda\right)}\mathcal{G}\left(p_{\Lambda}\right)\label{eq:Parisi}
\end{equation}
We obtained the variational bound for the quenched pressure, in particular,
we proposed in \cite{Franchini2024} that the upper bound of Eq. (\ref{eq:Parisi})
is analogue to the Parisi variational principle for the SK model \cite{Guerra_Bound,ASS,From_Parisi_to_Boltzmann}.
See Sections 2.b and 3.a of \cite{Franchini2024} for more details.

\begin{figure}
\begin{centering}
\includegraphics[scale=0.5]{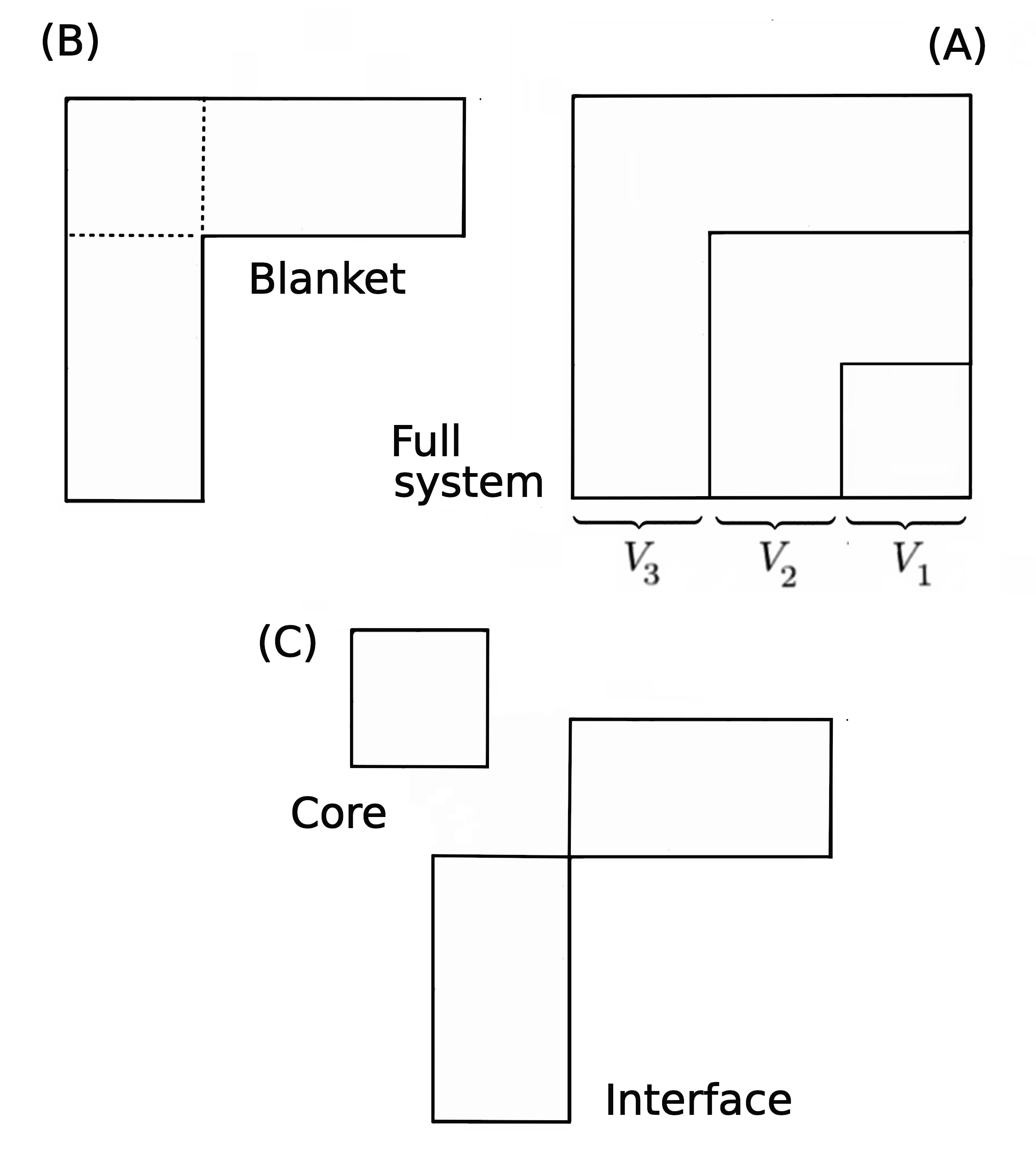}
\par\end{centering}
\begin{centering}
~
\par\end{centering}
\centering{}~\caption{\protect\label{fig:Blanket-decomposition-for-1}Blanket decomposition
of $W$ for two--body interactions $n=2$ \cite{RSBwR,Franchini2021}.}
\end{figure}

\section{Pairwise interactions}

Before describing how the Core--Interface picture of \cite{RSBwR,Franchini2021,FranchiniSPA2023,Franchini2024,Franchini_Pan_2025}
applies for the many body interactions, it will be useful to first
recall from \cite{Franchini2024} how the blankets look like for the
pairwise interactions. Let introduce the pairwise Hamiltonian
\begin{equation}
H^{\left(2\right)}\left(\sigma_{V}\right):=\sum_{i\in V}\sum_{j\in V}H_{ij}^{\left(2\right)}\sigma_{i}\sigma_{j}
\end{equation}
For two body interactions the coupling tensor reduces to a matrix.
Therefore, the edges set associated to the blanket is composed by
those edges with all the ends in $Q_{\ell}$ minus those with all
ends in $Q_{\ell-1}$, \cite{Franchini2024}. Formally:
\begin{equation}
W_{\ell}^{(2)}:=Q_{\ell}^{2}\setminus Q_{\ell-1}^{2}
\end{equation}
the number of edges inside $W_{\ell}$ is given by the formula
\begin{equation}
|W_{\ell}^{(2)}|/N^{2}=q_{\ell}^{2}-q_{\ell-1}^{2}
\end{equation}
To compute the Core--Interface picture of \cite{RSBwR,Franchini2021,FranchiniSPA2023,Franchini2024}
let us start from the identity:
\begin{multline}
H^{\left(2\right)}\left(\sigma_{Q_{\ell}}\right)=\sum_{i\in V_{\ell}}\sum_{j\in V_{\ell}}H_{ij}^{\left(2\right)}\sigma_{i}\sigma_{j}+\\
+\sum_{i\in V_{\ell}}\sum_{j\in Q_{\ell-1}}H_{ij}^{\left(2\right)}\sigma_{i}\sigma_{j}+\sum_{i\in Q_{\ell-1}}\sum_{j\in V_{\ell}}H_{ij}^{\left(2\right)}\sigma_{i}\sigma_{j}+H^{\left(2\right)}\left(\sigma_{Q_{\ell-1}}\right)\label{eq:fgbszgsz}
\end{multline}
Following \cite{RSBwR,Franchini2021,FranchiniSPA2023,Franchini2024},
we identify the Core of the blanket as the set of those edges with
all ends in $V_{\ell}$, while the Interface comprises edges with
one ends in $V_{\ell}$ and the other in $Q_{\ell-1}$. We immediately
identify the Core in
\begin{equation}
\sum_{i\in V_{\ell}}\sum_{j\in V_{\ell}}H_{ij}^{\left(2\right)}\sigma_{i}\sigma_{j}=H^{\left(2\right)}\left(\sigma_{V_{\ell}}\right)
\end{equation}
Let us introduce the reduced interaction matrix:
\begin{equation}
\bar{H}_{ij}^{\left(2\right)}:=H_{ij}^{\left(2\right)}+H_{j\,i}^{\left(2\right)}
\end{equation}
and the associated cavity field:
\begin{equation}
\bar{h}_{i}^{\left(2\right)}\left(\sigma_{Q_{\ell-1}}\right):=\sum_{j\in Q_{\ell-1}}\bar{H}_{ij}^{\left(2\right)}\sigma_{j}
\end{equation}
Then, the Interface is found through the following chain of equalities
\begin{multline}
\sum_{i\in V_{\ell}}\sum_{j\in Q_{\ell-1}}H_{ij}^{\left(2\right)}\sigma_{i}\sigma_{j}+\sum_{i\in Q_{\ell-1}}\sum_{j\in V_{\ell}}H_{ij}^{\left(2\right)}\sigma_{i}\sigma_{j}=\\
=\sum_{i\in V_{\ell}}\sum_{j\in Q_{\ell-1}}(\,H_{ij}^{\left(2\right)}+H_{ji}^{\left(2\right)})\,\sigma_{i}\sigma_{j}=\sum_{i\in V_{\ell}}\sigma_{i}\,\bar{h}_{i}^{\left(2\right)}\left(\sigma_{Q_{\ell-1}}\right)\label{eq:adfgarga}
\end{multline}
Putting together, we can rewrite the starting identity as
\begin{equation}
H^{\left(2\right)}\left(\sigma_{Q_{\ell}}\right)=H^{\left(2\right)}\left(\sigma_{V_{\ell}}\right)+\sum_{i\in V_{\ell}}\sigma_{i}\,\bar{h}_{i}^{\left(2\right)}\left(\sigma_{Q_{\ell-1}}\right)+H^{\left(2\right)}\left(\sigma_{Q_{\ell-1}}\right)
\end{equation}
Then, in the Blanket decomposition for $n=2$ the Lagrangian is \cite{RSBwR,Franchini2021,FranchiniSPA2023,Franchini2024}
\begin{equation}
\mathcal{L}_{\ell}\left(\sigma_{Q_{\ell}}\right)=\mathcal{C}_{\mathrm{\ell}}\left(\sigma_{V_{\ell}}\right)+\mathcal{I}_{\ell}\left(\sigma_{Q_{\ell}}\right)
\end{equation}
where $\mathcal{C}_{\mathrm{\ell}}$ and $\mathcal{I}_{\ell}$ are
the Core and the Interface respectively 
\begin{equation}
\mathcal{C}_{\mathrm{\ell}}\left(\sigma_{V_{\ell}}\right)=H^{\left(2\right)}\left(\sigma_{V_{\ell}}\right),\ \ \ \mathcal{I}_{\ell}\left(\sigma_{Q_{\ell}}\right)=\sum_{i\in V_{\ell}}\sigma_{i}\,\bar{h}_{i}^{\left(2\right)}\left(\sigma_{Q_{\ell-1}}\right)
\end{equation}
Therefore, the partition function of the Interface model for $n=2$
is
\begin{multline}
\bar{Z}_{\ell}\left(\sigma_{Q_{\ell-1}}\right)=\sum_{\sigma_{\,V_{\ell}}\in\left\{ -1,1\right\} ^{V_{\ell}}}\exp\mathcal{I}_{\ell}\left(\sigma_{Q_{\ell}}\right)=\\
=\sum_{\sigma_{\,V_{\ell}}\in\left\{ -1,1\right\} ^{V_{\ell}}}\exp\,\sum_{i\in V_{\ell}}\sigma_{i}\,\bar{h}_{i}^{\left(2\right)}\left(\sigma_{Q_{\ell-1}}\right)=\prod_{i\in V_{\ell}}2\cosh\bar{h}_{i}^{\left(2\right)}\left(\sigma_{Q_{\ell-1}}\right)\label{eq:fee-1}
\end{multline}
Taking the logarithm and normalizing
\begin{equation}
\bar{f}_{\ell}\left(\sigma_{Q_{\ell-1}}\right)=\frac{1}{\left|V_{\ell}\right|}\sum_{i\in V_{\ell}}\log2\cosh\bar{h}_{i}^{\left(2\right)}\left(\sigma_{Q_{\ell-1}}\right)
\end{equation}
we find the the functional to which apply the cascade of conditional
expectations of the previous section \cite{RSBwR,Franchini2021,FranchiniSPA2023,Franchini2024}. 

\begin{figure}
\begin{centering}
\includegraphics[scale=0.5]{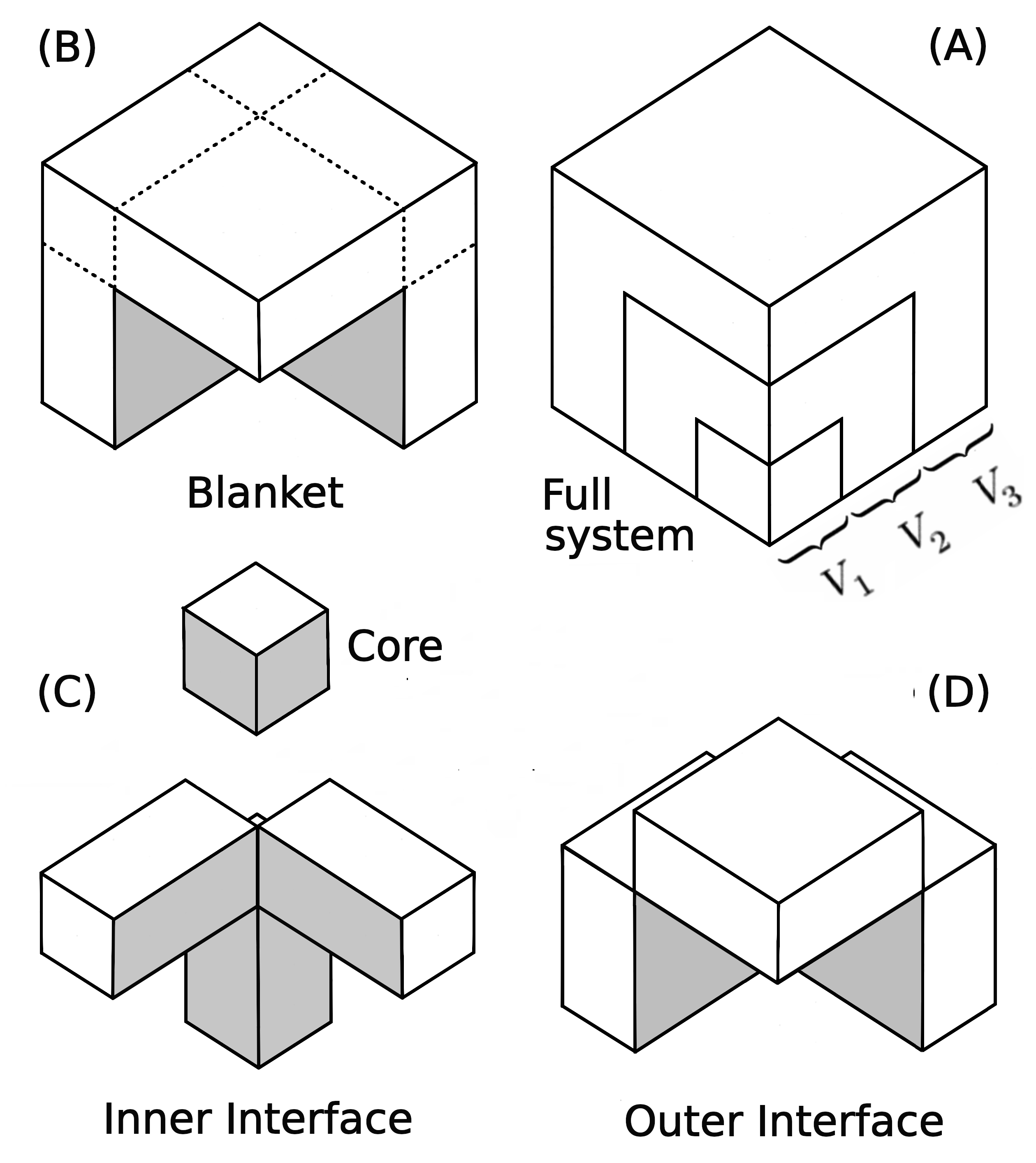}
\par\end{centering}
\begin{centering}
~
\par\end{centering}
\begin{centering}
~
\par\end{centering}
\centering{}\caption{\protect\label{fig:Blanket-decomposition-for}Blanket decomposition
of $W$ for three--body interactions $n=3$}
\end{figure}

\section{Three--body interactions}

As is anticipated in \cite{RSBwR}, the extension to $n-$body interactions
just require to consider an $n-$dimensional tensor on behalf of the
matrix ($n=2$) of Figure \ref{fig:Blanket-decomposition-for-1}.
The Core--Interface picture and the whole mathematical apparatus
is preserved but in this case the interface can be further destructured.
For simplicity here we will explicitly show only the case of three--body
interactions, that is easily visualized like in Figure \ref{fig:Blanket-decomposition-for},
but the generalization to $n-$body is straightforward. We start from
the three--body Hamiltonian
\begin{equation}
H^{\left(3\right)}\left(\sigma_{V}\right):=\sum_{i\in V}\sum_{j\in V}\sum_{k\in V}H_{ij\,k}^{\left(3\right)}\,\sigma_{i}\,\sigma_{j}\sigma_{k}
\end{equation}
By the definitions, the edges set of the $\ell-$th blanket is
\begin{equation}
W_{\ell}^{(3)}:=Q_{\ell}^{3}\setminus Q_{\ell-1}^{3}
\end{equation}
The number of edges that composes the blanket is therefore 
\begin{equation}
|W_{\ell}^{(3)}|/N^{3}=q_{\ell}^{3}-q_{\ell-1}^{3}
\end{equation}
As before, the Core--Interface picture \cite{RSBwR,Franchini2021,FranchiniSPA2023,Franchini2024}
is deduced from the identity
\begin{multline}
H^{\left(3\right)}\left(\sigma_{Q_{\ell}}\right)=\sum_{i\in V_{\ell}}\sum_{j\in V_{\ell}}\sum_{k\in V_{\ell}}H_{ij\,k}^{\left(3\right)}\sigma_{i}\sigma_{j}\sigma_{k}+\\
+\sum_{j\in V_{\ell}}\sum_{k\in V_{\ell}}\sigma_{j}\sigma_{k}\sum_{i\in Q_{\ell-1}}H_{ij\,k}^{\left(3\right)}\sigma_{i}+\sum_{i\in V_{\ell}}\sum_{k\in V_{\ell}}\sigma_{i}\sigma_{k}\sum_{j\in Q_{\ell-1}}H_{ij\,k}^{\left(3\right)}\sigma_{j}+\\
+\sum_{i\in V_{\ell}}\sum_{j\in V_{\ell}}\sigma_{i}\sigma_{j}\sum_{k\in Q_{\ell-1}}H_{ij\,k}^{\left(3\right)}\sigma_{k}+\\
+\sum_{i\in V_{\ell}}\sigma_{i}\sum_{j\in Q_{\ell-1}}\sum_{k\in Q_{\ell-1}}H_{ij\,k}^{\left(3\right)}\sigma_{j}\sigma_{k}+\sum_{j\in V_{\ell}}\sigma_{j}\sum_{i\in Q_{\ell-1}}\sum_{k\in Q_{\ell-1}}H_{ij\,k}^{\left(3\right)}\sigma_{i}\sigma_{k}+\\
+\sum_{k\in V_{\ell}}\sigma_{k}\sum_{i\in Q_{\ell-1}}\sum_{j\in Q_{\ell-1}}H_{ij\,k}^{\left(3\right)}\sigma_{i}\sigma_{j}+H^{\left(3\right)}\left(\sigma_{Q_{\ell-1}}\right)\label{eq:thsht}
\end{multline}
Notice that although the Core is essentially the same of the two--body
case, i.e.
\begin{equation}
\sum_{i\in V_{\ell}}\sum_{j\in V_{\ell}}\sum_{k\in V_{\ell}}H_{ijk}^{\left(3\right)}\sigma_{i\,}\sigma_{j}\sigma_{k}=H^{\left(3\right)}\left(\sigma_{V_{\ell}}\right)
\end{equation}
the Interface is more complex at first look, as we can immediately
identify at least two sub--components, shown in the Figure \ref{fig:Blanket-decomposition-for},
that we call Inner and Outer Interface. Let us introduce the auxiliary
notation
\begin{equation}
\bar{H}_{ij\,k}^{\left(3\right)}:=H_{ij\,k}^{\left(3\right)}+H_{j\,i\,k}^{\left(3\right)}+H_{kj\,i}^{\left(3\right)}\label{eq:MIX}
\end{equation}
and the re--normalized couplings
\begin{equation}
\bar{H}_{j\,k}^{\left(3\right)}\left(\sigma_{Q_{\ell-1}}\right):=\sum_{i\in Q_{\ell-1}}\sigma_{i}\,\bar{H}_{ij\,k}^{\left(3\right)},\ \ \ \bar{h}_{i}^{\left(3\right)}\left(\sigma_{Q_{\ell-1}}\right):=\sum_{j\in Q_{\ell-1}}\sum_{k\in Q_{\ell-1}}\bar{H}_{ij\,k}^{\left(3\right)}\sigma_{j}\sigma_{k}
\end{equation}
Then, the Hamiltonian associated to the Inner Interface is
\begin{multline}
\sum_{j\in V_{\ell}}\sum_{k\in V_{\ell}}\sigma_{j}\sigma_{k}\sum_{i\in Q_{\ell-1}}H_{ij\,k}^{\left(3\right)}\sigma_{i}+\sum_{i\in V_{\ell}}\sum_{k\in V_{\ell}}\sigma_{i}\sigma_{k}\sum_{j\in Q_{\ell-1}}H_{ij\,k}^{\left(3\right)}\sigma_{j}+\\
+\sum_{i\in V_{\ell}}\sum_{j\in V_{\ell}}\sigma_{i}\sigma_{j}\sum_{k\in Q_{\ell-1}}H_{ij\,k}^{\left(3\right)}\sigma_{k}=\\
=\sum_{j\in V_{\ell}}\sum_{k\in V_{\ell}}\sigma_{j}\sigma_{k}\sum_{i\in Q_{\ell-1}}(\,H_{ij\,k}^{\left(3\right)}+H_{j\,i\,k}^{\left(3\right)}+H_{kj\,i}^{\left(3\right)}\,)\,\sigma_{i}=\sum_{j\in V_{\ell}}\sum_{k\in V_{\ell}}\sigma_{j}\sigma_{k}\,\bar{H}_{j\,k}^{\left(3\right)}\left(\sigma_{Q_{\ell-1}}\right)\label{eq:dvfgs}
\end{multline}
and that associated to the Outer Interface is
\begin{multline}
\sum_{i\in V_{\ell}}\sigma_{i}\sum_{j\in Q_{\ell-1}}\sum_{k\in Q_{\ell-1}}H_{ij\,k}^{\left(3\right)}\sigma_{j}\sigma_{k}+\sum_{j\in V_{\ell}}\sigma_{j}\sum_{i\in Q_{\ell-1}}\sum_{k\in Q_{\ell-1}}H_{ij\,k}^{\left(3\right)}\sigma_{i}\sigma_{k}+\\
+\sum_{k\in V_{\ell}}\sigma_{k}\sum_{i\in Q_{\ell-1}}\sum_{j\in Q_{\ell-1}}H_{ij\,k}^{\left(3\right)}\sigma_{i}\sigma_{j}=\\
=\sum_{i\in V_{\ell}}\sigma_{i}\sum_{j\in Q_{\ell-1}}\sum_{k\in Q_{\ell-1}}(\,H_{ij\,k}^{\left(3\right)}+H_{j\,i\,k}^{\left(3\right)}+H_{kj\,i}^{\left(3\right)})\,\sigma_{j}\sigma_{k}=\sum_{i\in V_{\ell}}\sigma_{i}\,\bar{h}_{i}^{\left(3\right)}\left(\sigma_{Q_{\ell-1}}\right)\label{eq:cxhx}
\end{multline}
Combining together, we can rewrite the starting identity in the following
way 
\begin{multline}
H^{\left(3\right)}\left(\sigma_{Q_{\ell}}\right)=H^{\left(3\right)}\left(\sigma_{V_{\ell}}\right)+\sum_{j\in V_{\ell}}\sum_{k\in V_{\ell}}\sigma_{j}\sigma_{k}\,\bar{H}_{j\,k}^{\left(3\right)}\left(\sigma_{Q_{\ell-1}}\right)+\\
+\sum_{i\in V_{\ell}}\sigma_{i}\,\bar{h}_{i}^{\left(3\right)}\left(\sigma_{Q_{\ell-1}}\right)+H^{\left(3\right)}\left(\sigma_{Q_{\ell-1}}\right)\label{eq:vcsdvf}
\end{multline}
Therefore, the Blanket decomposition for $n=3$ is as follows: 
\begin{equation}
\mathcal{L}_{\ell}\left(\sigma_{Q_{\ell}}\right)=\mathcal{C}_{\mathrm{\ell}}\left(\sigma_{V_{\ell}}\right)+\mathcal{I}_{\ell}^{\left(\mathrm{in}\right)}\left(\sigma_{Q_{\ell}}\right)+\mathcal{I}_{\ell}^{\left(\mathrm{out}\right)}\left(\sigma_{Q_{\ell}}\right)
\end{equation}
where the core is like in the two--body case
\begin{equation}
\mathcal{C}_{\mathrm{\ell}}\left(\sigma_{V_{\ell}}\right)=H^{\left(3\right)}\left(\sigma_{V_{\ell}}\right)
\end{equation}
Then we have an Inner Interface
\begin{equation}
\mathcal{I}_{\ell}^{\left(\mathrm{in}\right)}\left(\sigma_{Q_{\ell}}\right)=\sum_{j\in V_{\ell}}\sum_{k\in V_{\ell}}\sigma_{j}\sigma_{k}\,\bar{H}_{j\,k}^{\left(3\right)}\left(\sigma_{Q_{\ell-1}}\right)
\end{equation}
that is two--body Hamiltonian, and the Outer Interface 
\begin{equation}
\mathcal{I}_{\ell}^{\left(\mathrm{out}\right)}\left(\sigma_{Q_{\ell}}\right)=\sum_{i\in V_{\ell}}\sigma_{i}\,\bar{h}_{i}^{\left(3\right)}\left(\sigma_{Q_{\ell-1}}\right)
\end{equation}
that is like before, a one--body Hamiltonian. Remarkably, from Figure
\ref{fig:Blanket-decomposition-for} is quite clear that in fully
connected models the contribution from the outer interface will overwhelm
both the Core and the Inner Interface in the limit of infinite blankets,
therefore, we may construct the IO model of \cite{RSBwR} from this
component only. The IO partition function is 
\begin{multline}
\bar{Z}_{\ell}\left(\sigma_{Q_{\ell-1}}\right)=\sum_{\sigma_{\,V_{\ell}}\in\left\{ -1,1\right\} ^{V_{\ell}}}\exp\mathcal{I}_{\ell}^{\left(\mathrm{out}\right)}\left(\sigma_{Q_{\ell}}\right)=\\
=\sum_{\sigma_{\,V_{\ell}}\in\left\{ -1,1\right\} ^{V_{\ell}}}\exp\,\sum_{i\in V_{\ell}}\sigma_{i}\,\bar{h}_{i}^{\left(3\right)}\left(\sigma_{Q_{\ell-1}}\right)=\prod_{i\in V_{\ell}}2\cosh\bar{h}_{i}^{\left(3\right)}\left(\sigma_{Q_{\ell-1}}\right)\label{eq:fee}
\end{multline}
and the corresponding pressure density is
\begin{equation}
\bar{f}_{\ell}\left(\sigma_{Q_{\ell-1}}\right)=\frac{1}{\left|V_{\ell}\right|}\sum_{i\in V_{\ell}}\log2\cosh\bar{h}_{i}^{\left(3\right)}\left(\sigma_{Q_{\ell-1}}\right)
\end{equation}
We immediately appreciate that although the differences in the coefficients,
the general picture is formally the same of the two--body case, at
least in the limit of infinitely thin blankets. See \cite{Franchini2024,Franchini_Pan_2025}
for how to proceed from this point or \cite{RSBwR,Franchini2021,FranchiniSPA2023}
for more details.

\subsection{Example: Constructive Cavity Method for $n=3$}

To conclude, we will show how to combine the above findings with the
\textbf{Cavity Method} \cite{Parisi_Mezard_Virasoro,ASS} to find
the Random Overlap Structure (ROSt) \cite{ASS} of the multi--spin
SK model in a constructive way \cite{RSBwR,Franchini2021,FranchiniSPA2023,FraConcetti}.
Let us consider i.i.d. Gaussian interactions
\begin{equation}
H_{ij\,k}^{\left(3\right)}=\beta J_{ij\,k}/N,\ \ \ J_{ij\,k}\sim\mathscr{N}\left(0,1\right)
\end{equation}
By previous considerations we can generalize the arguments of \cite{ASS}
and find the cavity functional for the $3-$spin SK model:
\begin{equation}
A\left(\mu\right):=\log\langle2\cosh\,a^{\left(3\right)}h_{N+1}^{\left(3\right)}\left(\sigma_{V}\right)\rangle_{\mu^{*}}-\log\langle\exp\,b_{N}^{\left(3\right)}H^{\left(3\right)}\left(\sigma_{V}\right)\rangle_{\mu^{*}}
\end{equation}
The coefficient $a^{\left(3\right)}$ is due to the to the Gaussian
summation rule applied to the Eq. (\ref{eq:MIX}) and can be deduced
directly by confronting with the distributional equation
\begin{equation}
\bar{H}_{ij\,k}^{\left(3\right)}\stackrel{d}{=}a^{\left(3\right)}H_{ij\,k}^{\left(3\right)}
\end{equation}
For $n=3$ we therefore find the following:
\begin{equation}
a^{\left(3\right)}=\sqrt{3}
\end{equation}
In a general $n-$body interaction we would have found a multiplicity
$n$. The coefficient of the Hamiltonian in the correction term also
can be computed (by another clever use of the Gaussian summation rule
\cite{ASS}). We Taylor expand the square of the normalization of
the system with $N+1$ spins: for $3-$body interactions 
\begin{equation}
\left(N+1\right)^{-2}=\frac{1}{N^{2}}+\frac{2}{N^{3}}+\,...\,
\end{equation}
By Gaussian summation rule, the coefficient $b_{N}^{\left(3\right)}$
is the square root of the second term of this expansion, therefore
we deduce the following value
\begin{equation}
b_{N}^{\left(3\right)}=\sqrt{\frac{2}{N^{3}}}
\end{equation}
In $n-$body interactions we would find the coefficient $n-1$ instead
of $2$ due to the $n-1$ on top of the second term of the expansion,
and an exponent $n$ instead of $3$. 

Let now construct the ROSt variables: the first is obtained by applying
the partition in Figure \ref{fig:Blanket-decomposition-for-1} to
the cavity field. By introducing the set of normalized Gaussian variables
\begin{equation}
x_{\ell}(\sigma_{Q_{\ell}}):=\frac{1}{\sqrt{|W_{\ell}^{(2)}|}}\sum_{j\,k\in W_{\ell}^{(2)}}J_{N+1j\,k}\,\sigma_{j}\sigma_{k}
\end{equation}
and recalling that the size (cardinality) of $W_{\ell}^{(2)}$ is
\begin{equation}
|W_{\ell}^{(2)}|=(q_{\ell}^{2}-q_{\ell-1}^{2})\,N^{2}
\end{equation}
we find the first expression: 
\begin{equation}
a^{\left(3\right)}h_{N+1}^{\left(3\right)}\left(\sigma_{V}\right)=\beta\sum_{\ell\leq L}x_{\ell}(\sigma_{Q_{\ell}})\sqrt{3\,(q_{\ell}^{2}-q_{\ell-1}^{2})}
\end{equation}
For the correction term we must use the partition shown in Figure
\ref{fig:Blanket-decomposition-for} instead: let introduce another
set of normalized variables
\begin{equation}
y_{\ell}(\sigma_{Q_{\ell}}):=\frac{1}{\sqrt{|W_{\ell}^{(3)}|}}\sum_{ij\,k\in W_{\ell}^{(3)}}J_{ij\,k}\,\sigma_{\,i}\sigma_{j}\sigma_{k}
\end{equation}
and recall that the size of the set $W_{\ell}^{(3)}$ is 
\begin{equation}
|W_{\ell}^{(3)}|=(q_{\ell}^{3}-q_{\ell-1}^{3})\,N^{3}
\end{equation}
Then, it is easy to verify that the following holds:
\begin{equation}
b_{N}^{\left(3\right)}H^{\left(3\right)}\left(\sigma_{V}\right)=\beta\sum_{\ell\leq L}y_{\ell}(\sigma_{Q_{\ell}})\sqrt{2\,(q_{\ell}^{3}-q_{\ell-1}^{3})}
\end{equation}
The final expression for the functional is therefore 
\begin{multline}
A\left(\mu\right)=\log\langle2\cosh\,\beta\sum_{\ell\leq L}x_{\ell}(\sigma_{Q_{\ell}})\sqrt{3\,(q_{\ell}^{2}-q_{\ell-1}^{2})}\rangle_{\mu^{*}}+\\
-\log\langle\exp\,\beta\sum_{\ell\leq L}y_{\ell}(\sigma_{Q_{\ell}})\sqrt{2\,(q_{\ell}^{3}-q_{\ell-1}^{3})}\rangle_{\mu^{*}}\label{eq:dfdgdgdfgdfgdfd}
\end{multline}
Finally, to confront with the formulas in \cite{Talagrand} we have
to apply the following re--scaling of the temperature due to considering
the asymmetric version: 
\begin{equation}
\beta\rightarrow\beta/\sqrt{6}
\end{equation}
In fact, notice that in the definitions of \cite{Talagrand} the permutations
are considered only once. To implement the same constraint in the
$n-$body case the coefficient inside the square root should be $n!$.
It is now evident by comparison with \cite{Talagrand} that the variables
$x_{\ell}$ and $y_{\ell}$ are the ROSt variables of \cite{ASS},
and that computing the averages according to a Ruelle Cascade \cite{FraConcetti}
will end up in the Parisi functional as defined in \cite{Talagrand}. 

We remark that this strongly supports the Blanket picture given before,
since to the best of our knowledge no construction of the ROSt variables
has ever been proposed so far, either by standard methods or by any
other method, except in \cite{RSBwR,Franchini2021,FraConcetti}.

\section{Acknowledgments }

This project received fundings from the European Research Council
(ERC), under European Union\textquoteright s Horizon 2020 Research
and Innovation programme, Grant agreement No. {[}694925{]}.

\end{document}